\documentclass[sigconf,nonacm]{acmart}

\AtBeginDocument{%
  }

\usepackage{makecell}
\usepackage[table]{xcolor}

\setcopyright{none}
\renewcommand{\footnotetextcopyrightpermission}[1]{}
\begin{document}

\title{Design and Evaluation of a Culturally Adapted Multimodal Virtual Agent for PTSD Screening}

\author{Cengiz Ozel}
\affiliation{%
  \institution{Ministry of Defense}
  \city{Riyadh}
  \country{Saudi Arabia}
}
\email{cozel@u.rochester.edu}

\author{Waleed Nadeem}
\authornote{Both authors contributed equally to this research.}
\author{Samuel Potter}
\authornotemark[1]
\affiliation{%
  \institution{Ministry of Defense}
  \country{United States of America}
}
\email{wnadeem@u.rochester.edu}
\email{s.potter@rochester.edu}

\author{Yahya Bokhari}
\affiliation{%
  \institution{Ministry of Defense}
  \city{Riyadh}
  \country{Saudi Arabia}
}
\email{yabokhari@mod.gov.sa}

\author{Bdour Alwuqaysi}
\affiliation{%
  \institution{Ministry of Defense}
  \city{Riyadh}
  \country{Saudi Arabia}
}
\email{balwuqaysi@mod.gov.sa}

\author{Wejdan Alotaibi}
\affiliation{%
  \institution{Prince Sultan Military Medical City}
  \city{Riyadh}
  \country{Saudi Arabia}
}
\email{wjdanotb@hotmail.com}

\author{Rahaf Fahad Alnufaie}
\affiliation{%
  \institution{Prince Sultan Military Medical City}
  \city{Riyadh}
  \country{Saudi Arabia}
}
\email{rahaffalnufaie@gmail.com}

\author{Sabri Boughorbel}
\affiliation{%
  \institution{Ministry of Defense}
  \city{Riyadh}
  \country{Saudi Arabia}
}
\email{sabri.boughorbel@gmail.com}

\author{Abdulrhman Aljouie}
\affiliation{%
  \institution{Ministry of Defense}
  \city{Riyadh}
  \country{Saudi Arabia}
}
\email{aljouie@gmail.com}

\author{Rakan Altasan}
\affiliation{%
  \institution{Prince Sultan Military Medical City}
  \city{Riyadh}
  \country{Saudi Arabia}
}
\email{rakanaltasan@gmail.com}

\author{Ehsan Hoque}
\affiliation{%
  \institution{Ministry of Defense}
  \city{Riyadh}
  \country{Saudi Arabia}
}
\email{mehoque@cs.rochester.edu}

\renewcommand{\shortauthors}{Ozel et al.}

\begin{abstract}
Post-traumatic stress disorder (PTSD) is highly prevalent yet chronically underreported among combat-exposed military personnel. This paper presents Molhim, a culturally adapted multimodal conversational AI platform that supports purpose-specific interactions through a configurable conversational pipeline consisting of session setup, real-time dialogue with a high-fidelity virtual avatar, and post-session analysis and feedback. In this work, we examine the PTSD screening configuration of the Molhim platform in a military healthcare context. The system employs a conversational avatar driven by a large language model, integrating real-time speech recognition, visual understanding of user input, text-to-speech synthesis, and a high-fidelity human avatar to support structured multi-turn dialogue and automated post-session analysis, including administration of the PTSD Checklist for DSM-5 (PCL-5). These findings suggest the feasibility of Molhim as a conversational platform for PTSD screening and highlight design considerations for socially cooperative human–AI systems in clinical environments.
\end{abstract}

\keywords{virtual agents, PTSD screening, military mental health, conversational AI, social cognition, human-AI cooperation, Saudi Arabia}

\maketitle

\section{Introduction}
PTSD is one of the most common and persistent mental health conditions affecting military personnel exposed to combat and deployment-related trauma \cite{steenkamp2015psychotherapy,lucas2017reporting}. Exposure to life-threatening events, prolonged operational stress, and reintegration challenges can significantly increase the likelihood of developing PTSD symptoms among soldiers and veterans \cite{polusny2011prospective}. Early identification of PTSD symptoms is critical, as timely screening and intervention can substantially improve long-term psychological outcomes and reduce the risk of chronic impairment \cite{megnin2019post}. However, despite the availability of validated clinical screening instruments, PTSD may remain underreported within military populations \cite{milliken2007longitudinal,rozek2023understanding}.

A major barrier to effective PTSD screening in military environments is the presence of social and institutional stigma surrounding mental health. Soldiers may fear that disclosing psychological distress could negatively affect their professional reputation, career progression, or relationships with peers and supervisors \cite{warner2008soldier,sharp2015stigma}. These concerns can lead individuals to avoid seeking help or delay engaging with mental health services \cite{hoge2004combat}. Consequently, researchers have increasingly explored alternative interaction approaches that reduce perceived judgment and encourage more open disclosure of sensitive psychological information \cite{croes2024digital,meng2021emotional,lee2025artificial}.

Conversational artificial intelligence has emerged as promising technologies for addressing these challenges. Virtual human interviewers can simulate elements of face-to-face conversation while providing a psychologically safe interaction space in which users may feel more comfortable discussing personal experiences \cite{devault2014simsensei}. Individuals interacting with virtual agents have been shown to disclose sensitive information more readily than in human interviews, a phenomenon referred to as the disclosure advantage \cite{lucas2014s}. These findings suggest that conversational agents may serve as scalable tools for early mental health screening, particularly when stigma and resource constraints limit access to traditional clinical care.

Deploying AI-driven mental health tools in real-world healthcare systems introduces challenges related to cultural adaptation, language support, and contextual alignment \cite{naderbagi2024cultural}. In regions such as the Middle East, conversational AI must account for language-specific design considerations \cite{fadhil2019ollobot}, and systems built for English-speaking populations may not transfer effectively to Arabic contexts without careful attention to dialect variation and diglossia \cite{farghaly2009arabic}.
To address these challenges, this paper introduces \textit{Molhim} (\textit{mulhim}, meaning "the Inspirer"), a multimodal conversational AI platform supporting purpose-specific interactions across domains. We examine a PTSD screening configuration deployed in a military healthcare context. Molhim integrates automatic speech recognition (ASR), large language models (LLMs), a vision–language model (VLM), and text-to-speech (TTS) synthesis to enable multi-turn dialogue through a human-like conversational interface. The platform is not designed to diagnose, treat, or replace clinical care. Instead, it functions as a structured screening and psychoeducational interface that guides users through evidence-based instruments such as the PCL-5, offers brief grounding exercises, and directs individuals toward professional resources when needed. The system employs a structured dialogue state machine to maintain conversational coherence and safety while supporting early identification and referral.

To evaluate the feasibility of this approach, we conducted a mixed-methods pilot study. In this study, returning soldiers observed therapist-guided demonstrations of the system and provided feedback on acceptability, safety, and perceived usefulness. Clinicians simultaneously documented observational reflections on participant comfort, system behavior, and potential clinical value.

This paper makes three primary contributions:

\begin{itemize}

\item \textbf{A multimodal conversational platform for PTSD screening.}
We present Molhim, a modular conversational AI system that integrates speech recognition, large language models, text-to-speech synthesis, visual context processing, and avatar-based interaction within a structured dialogue pipeline for mental health screening.

\item \textbf{A culturally adapted Arabic conversational screening design.}
We design a PTSD screening workflow tailored to Arabic-speaking military contexts, incorporating structured dialogue states, safety-aware interaction design, grounding strategies, and culturally aligned conversational behaviors.

\item \textbf{An exploratory feasibility evaluation in a military healthcare setting.}
We report findings from a mixed-methods pilot study conducted in a Saudi Ministry of Defense healthcare center, providing early evidence of system acceptability, perceived safety, and usefulness among soldiers and clinicians.

\end{itemize}

To further structure the exploratory evaluation of these contributions, we investigate the following research questions:
\begin{itemize}
    \item \textbf{RQ1:} Is the system perceived as safe and acceptable in a military healthcare context?
    \item \textbf{RQ2:} Does the system demonstrate culturally appropriate interaction?
    \item \textbf{RQ3:} Do clinicians consider the system usable in practice?
\end{itemize}

\section{Related Work}
\subsection{PTSD in Military Populations}
PTSD is a prevalent mental health condition among military personnel exposed to combat and deployment-related trauma \cite{armenta2018factors}. Combat exposure, operational stress, and reintegration challenges increase the risk of developing PTSD symptoms among soldiers and veterans \cite{hoge2004combat,polusny2011prospective}. Symptoms are often under-identified during early post-deployment screening, with many cases emerging months later \cite{milliken2007longitudinal}. Although validated instruments such as the PCL and PC-PTSD demonstrate strong reliability \cite{bliese2008validating}, challenges remain in achieving timely detection of PTSD symptoms in practice.

Underreporting is influenced by social and institutional stigma. Service members may avoid disclosing psychological distress due to concerns about career impact, unit perception, and professional reputation \cite{warner2008soldier,lucas2017reporting}. As a result, alternative screening approaches that reduce perceived judgment have been increasingly explored.

\subsection{Virtual Humans and the Disclosure Advantage}
Virtual human interviewers have emerged as a promising approach for psychological assessment, reducing perceived social judgment. Studies show that individuals may disclose more sensitive information to virtual agents than to human interviewers, a phenomenon known as the disclosure advantage \cite{lucas2014s}, including military populations reporting more PTSD symptoms to virtual interviewers \cite{lucas2017reporting}. However, conversational realism can also influence socially desirable responding, highlighting the need to balance rapport with truthful reporting \cite{schuetzler2018influence,pickard2016revealing}. These findings support virtual agents as interactive interfaces for social interaction and multimodal communication training \cite{hasan2023sapien,ali2017social}.

\subsection{Conversational Agents for PTSD Screening and Support}
A growing body of research has explored conversational agents designed for PTSD screening, psychoeducation, or self-management support. Mobile applications such as PTSD Coach demonstrate the potential of evidence-based digital tools for supporting individuals with PTSD \cite{kuhn2014preliminary,owen2015mhealth}. Building on such resources, the PTSDialogue system is a conversational agent that supports individuals through structured self-management dialogues adapted from PTSD Coach \cite{han2021ptsdialogue}. The system employs a finite-state dialogue structure intended to ensure safety and predictable conversational flow, similar to structured dialogue management approaches used in other virtual agent systems for healthcare communication and training \cite{kane2022flexible}. Subsequent evaluations of PTSDialogue included interviews with clinical experts to assess usability, perceived safety, and potential clinical value \cite{han2023preliminary}, as well as studies involving individuals living with PTSD to examine acceptability and user experience \cite{han2024assessing}.

Other work has explored virtual agents as tools supporting PTSD treatment processes rather than screening alone. For example, virtual agents have been used to support the reconstruction of traumatic memories through structured storytelling within digital environments \cite{tielman2017therapy}. While such systems demonstrate the potential role of conversational agents in therapeutic contexts, many remain small-scale prototypes primarily focused on treatment support. More broadly, conversational systems such as Woebot \cite{fitzpatrick2017delivering} and other dialogue-based interventions \cite{fulmer2018using} provide interactive support but are text-based and lack multimodal embodiment. Fully multimodal conversational systems combining voice interaction, avatar embodiment, and structured screening protocols remain relatively limited.

\subsection{Multimodal Behavioral Analysis and Automated PTSD Detection}
Automated detection of psychological distress using multimodal behavioral signals has been studied using datasets such as the Distress Analysis Interview Corpus (DAIC), which includes audio, video, transcripts, and questionnaire data collected during virtual agent interviews \cite{lucas2014s,gratch2014daic}. Prior work has developed models that infer mental health conditions from linguistic, acoustic, and visual features \cite{al2018detecting}, including approaches based on sentiment and language analysis of transcripts \cite{sawalha2022detecting}, speech-based acoustic modeling for PTSD detection and severity estimation \cite{hu2024speech}, and more recent LLM–based methods applied to clinical interview data \cite{chen2025detecting}. While these studies demonstrate the feasibility of automated PTSD detection, they are primarily based on offline analysis of recorded data, with limited exploration of integration into real-time interactive conversational screening systems.

\subsection{Interaction Quality in Conversational Systems}
Interaction quality is critical in conversational agents, shaping user perceptions of trust and engagement. Factors such as turn-taking, response latency, and conversational repair strongly influence perceived naturalness \cite{skantze2021turn}. In particular, delays between user input and system response can disrupt conversational flow, especially in speech-based interactions, and have been studied through strategies that use embodied or conversational cues to make system processing delays more transparent \cite{pelikan2023managing}. These effects are especially important in sensitive contexts such as mental health screening, where delays may reduce perceived empathy or engagement, yet remain underexplored in PTSD-focused conversational systems.

\subsection{Cultural and Linguistic Considerations in Arabic Conversational Systems}
Deploying conversational AI in Arabic-speaking contexts requires addressing dialect variation and diglossia, as everyday spoken Arabic differs significantly from Modern Standard Arabic. Conversational agents such as Nabiha and OlloBot demonstrate the importance of dialectal adaptation for accessibility and user engagement, particularly in health-related interactions \cite{al2020nabiha,fadhil2019ollobot}. Studies of AI adoption in Saudi healthcare further show that trust is shaped not only by technical performance but also by psychosocial and relational factors, including patient–doctor dynamics \cite{alharbi2026determinants}. Cross-linguistic research also indicates that conversational behaviors such as repair strategies vary across languages and cultural contexts \cite{alghamdi2025effect}.

\section{The Molhim Platform for PTSD Screening}
\begin{figure}[t]
\centering
\includegraphics[width=1.0\linewidth]{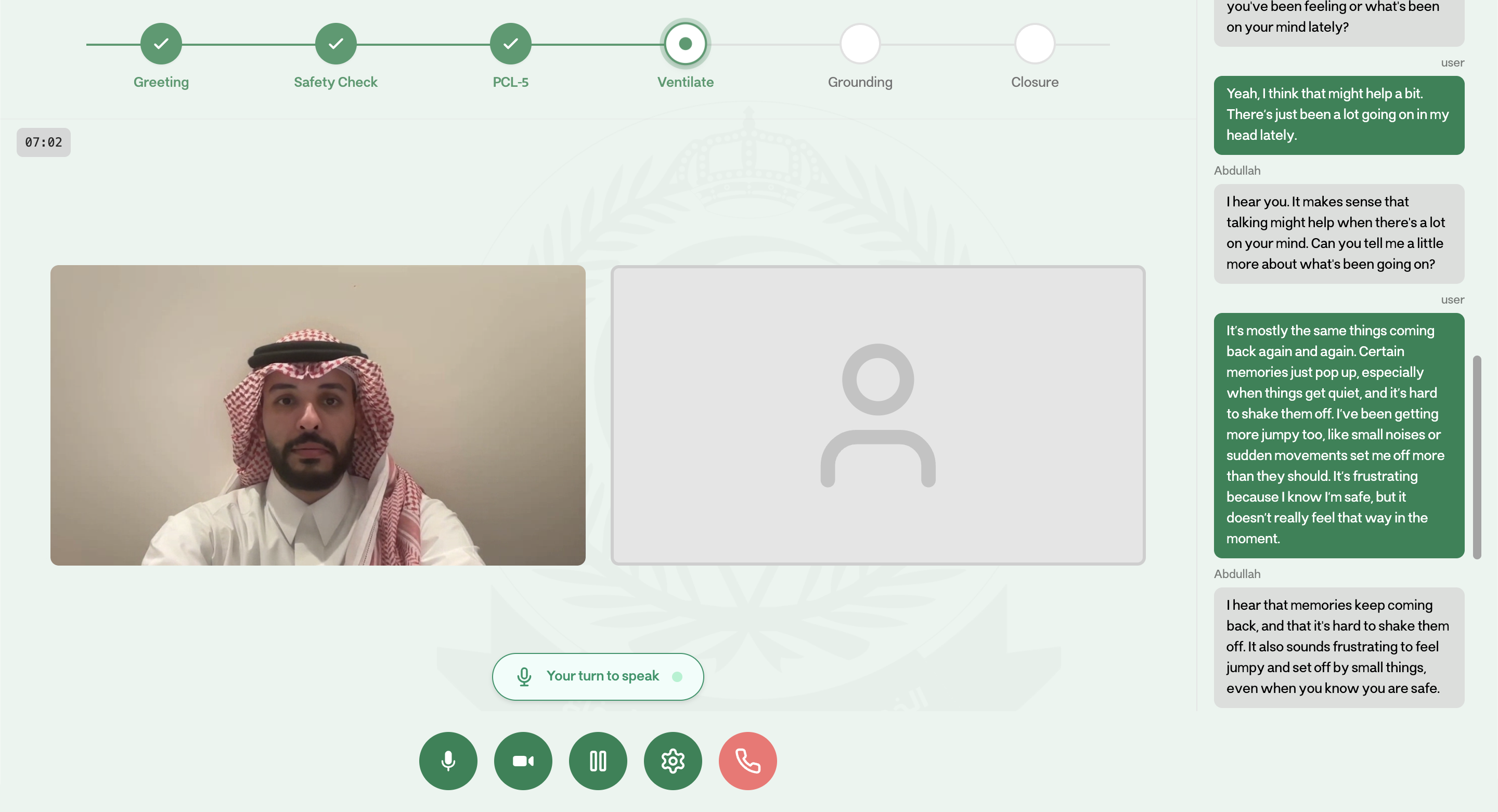}
\caption{Molhim conversational interface. Users interact with a virtual avatar while the system guides PTSD screening. The interface is localized for Saudi users, with an avatar in traditional attire to support cultural familiarity.}
\label{fig:meeting}
\end{figure}

\subsection{Architecture Overview}
\begin{figure*}[t]
\centering
\includegraphics[width=0.98\linewidth]{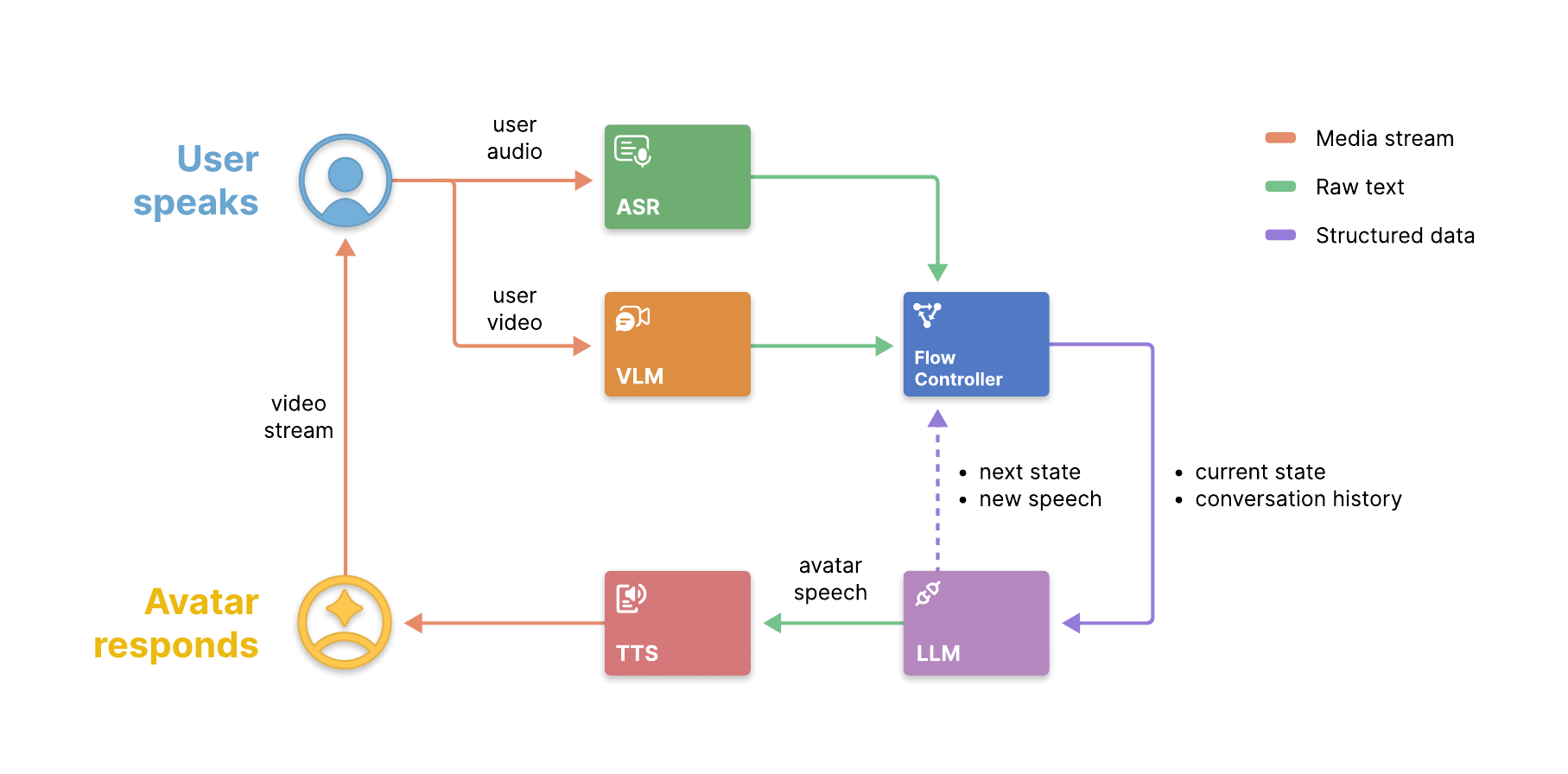}
\caption{System architecture of Molhim. User speech is transcribed by ASR and visual cues are extracted by a VLM. A flow controller manages the dialogue via a state machine and constructs prompts for the LLM, whose responses are synthesized to speech and delivered through a virtual avatar.}
\label{fig:architecture}
\end{figure*}

The platform is a modular multimodal system built around a unified interaction pipeline that combines state-based dialogue management with LLMs and avatar-based rendering. As illustrated in Figure~\ref{fig:architecture}, user speech is processed by an ASR component to produce a transcript, while a VLM extracts contextual cues from the webcam feed. These inputs are passed to a central flow controller that combines a state machine with LLM-driven response generation to orchestrate dialogue.

Speech recognition is implemented using an ASR system configured for Saudi Arabic with English code-switching support. For visual analysis, one webcam frame is captured per turn and processed by a VLM to extract coarse facial affect cues (e.g., general affect and engagement signals). This lightweight, per-turn sampling approach was selected to minimize computational overhead and avoid continuous video processing. The resulting description is used only to provide supplementary context for that turn’s response generation and is not relied upon for clinical inference or decision-making. Raw images are not stored or used in subsequent stages. This design reflects an exploratory use of visual signals in the current system, where multimodal inputs are intended to complement the conversational interaction rather than serve as a primary source of assessment.

Each use case is defined as a system prompt with a sequence of user-facing interaction stages and underlying dialogue states connected through explicit transition rules. Interaction stages provide a high-level structure visible to the user, while each stage may consist of multiple internal dialogue states that enable finer-grained control of the conversation. The controller constructs a per-turn prompt incorporating the current state, conversation summary, recent history, and visual cues. The LLM generates the avatar's response and selects the next state from a constrained set of allowable transitions, each associated with specific criteria. Additional constraints based on turn count and elapsed time further regulate progression, ensuring adherence to the intended screening protocol and preventing out-of-scope responses.

The generated response is converted to speech via TTS and delivered through a virtual avatar. Avatar rendering and real-time delivery are implemented using a real-time conversational video interface platform over WebRTC infrastructure, enabling synthesized speech to be streamed to a high-fidelity virtual avatar for lip-synced response delivery. The avatar was developed using a custom generation pipeline, in which a human actor wearing traditional Saudi attire was recorded to produce a culturally aligned virtual persona.

For dialogue generation and post-session analysis, the platform uses an LLM for conversational response generation and analysis tasks such as cognitive distortion identification and referral summary generation, while PCL-5 scoring is computed deterministically based on questionnaire responses.

\begin{figure}[t]
\centering
\includegraphics[width=1.0\linewidth]{Figures/post-session.png}
\caption{Example post-session results page. The interface summarizes the screening interaction, displaying the PCL-5 score (50/80), symptom group indicators, and a recommendation encouraging the user to seek professional support.}
\label{fig:postsession}
\end{figure}

\subsection{PTSD Screening Mode}
\begin{figure}[t]
\centering
\includegraphics[width=1\linewidth]{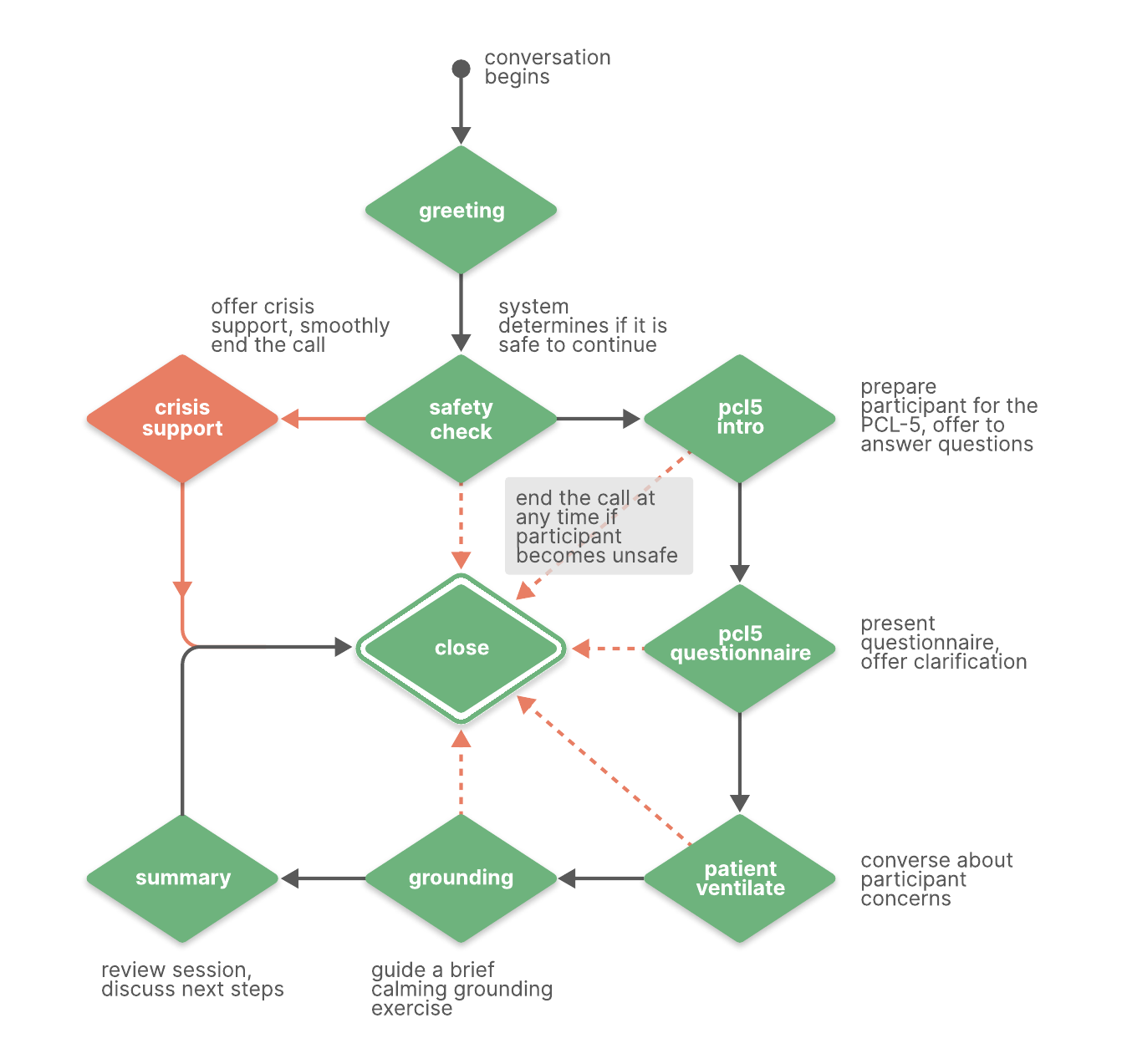}
\caption{State-machine structure of the Molhim PTSD screening interaction. Each node represents a conversational state with specific clinical objectives. Elevated risk during assessment triggers a diversion to a crisis-support state.}
\label{fig:ptsd-sm}
\end{figure}

Building on the architecture described above, the PTSD screening mode is implemented as a structured conversational protocol using a dialogue state machine, as illustrated in Figure~\ref{fig:ptsd-sm}. The interaction begins with a \textit{greeting} state in which the avatar introduces itself and establishes rapport with the user. The system then transitions to a \textit{safety check} stage that evaluates whether the user may be experiencing acute distress or crisis-related symptoms. If such concerns are identified, the flow diverts to a \textit{crisis support} state that prioritizes safety-oriented responses and directs the user toward appropriate support resources. If no immediate crisis signals are detected, the system proceeds to the \textit{PCL-5 introduction} state, where the user is informed about the PCL-5 and the purpose of the screening process. The conversation then transitions to the \textit{PCL-5 questionnaire} stage, during which the system administers the structured assessment.

\begin{table}[t]
\centering
\caption{Dialogue states and prompt summaries used in the PTSD screening workflow}
\label{tab:dialogue_states}
\rowcolors{2}{gray!20}{white}
\begin{tabular}{p{1.5cm}p{6.3cm}}
\hline
\textbf{State} & \textbf{Prompt Summary} \\
\hline

\texttt{greeting} & 
Warmly greet the user and ask how they are feeling today using an open-ended question. \\

\texttt{safety check} & 
Assess user’s emotional state and confirm safety. Check for immediate risk if distress is detected. \\

\texttt{crisis support} & 
Prioritize the user’s safety with calm, supportive language. Encourage contacting emergency services or a trusted person. \\

\texttt{pcl5 intro} & 
Explain the purpose of the PCL-5 and its time frame. Inform the user that a questionnaire will appear and outline next steps. \\

\texttt{\makecell[tl]{pcl5\\question-\\naire}} &
Notify the user that the questionnaire is visible. Encourage them to take their time and offer clarification if needed. \\

\texttt{patient ventilate} & 
Invite the user to share their experiences and current feelings through open-ended dialogue. Reflect responses, validate emotions, and support expression using gentle prompts if needed. \\

\texttt{grounding} & 
Guide the user through a brief grounding exercise focused on sensory awareness. Maintain a calm, optional tone to support emotional stabilization. \\

\texttt{summary} & 
Provide a brief, non-diagnostic synthesis of key themes from the interaction. Suggest one or two supportive next steps. \\

\texttt{close} & 
Thank user for sharing and reinforce support availability. Conclude session with a respectful sign-off. \\

\texttt{end} & 
Politely terminate the session without introducing new content. Final state of the interaction. \\

\hline
\end{tabular}
\end{table}

Following completion of the questionnaire, the flow enters a \textit{patient ventilation} stage that allows the user to elaborate on their experiences in a more open-ended conversational format. This is followed by a \textit{grounding} stage that provides brief coping strategies or stabilization exercises designed to support emotional regulation after discussing potentially distressing topics. Finally, the interaction transitions to a \textit{summary} state in which the system summarizes key aspects of the conversation before moving to the \textit{close} state that concludes the session. At any point during the interaction, the user may end the session either by terminating the call or by indicating that they would like to do so. In such cases, the agent acknowledges the request and transitions smoothly to the closing state.

After the session, the system generates a structured post-session report summarizing screening outcomes and interaction insights, including PCL-5 scoring with symptom cluster breakdowns, LLM-generated recommendations with reasoning, and cognitive distortion analysis accompanied by brief psychoeducational resources, as illustrated in Figure~\ref{fig:postsession}.

\subsection{Clinical Co-Design and Iterative Development}
The PTSD screening configuration of Molhim was developed through ongoing iteration with licensed mental health professionals, with repeated refinements driven by clinical review of the conversational flow, session structure, and safety mechanisms. Rather than defining the screening experience solely from a technical perspective, the system was shaped to better reflect how therapists felt a trauma-informed screening interaction should unfold in practice.

Early versions of the workflow emphasized the technical foundations of multimodal conversation and structured state-based dialogue. Through repeated feedback from clinicians, the interaction was progressively revised to better align with real screening needs and with the emotional realities of discussing trauma-related symptoms. This process led to the incorporation of grounding exercises, more careful handling of sensitive content, and clearer boundaries around the role of the system as a screening and referral tool rather than a therapeutic intervention.

The flow was updated to support emergency-stop cues, allow users to skip grounding or open-ended ventilation stages, and provide a natural option to end the session after completion of the PCL-5 rather than forcing continued interaction. The framing of certain stages was also revised to better match trauma-informed practice, including the shift from more evaluative language such as ``risk assessment'' toward softer and more user-centered framing such as ``safety check.'' These changes helped move the system away from a rigid scripted protocol and toward a more clinically appropriate and user-sensitive screening experience.

Therapist feedback also informed workflow decisions beyond a single session. As the platform evolved, features were added to support private mode for sensitive sessions, in which interactions are not stored or retained after completion, alongside consent-oriented interaction design and follow-up flows for returning users so repeated sessions could build on prior context rather than re-administering the same screening content. The iterative development process helped ensure the platform was technically functional and better aligned with therapist expectations regarding safety, pacing, flexibility, and referral-oriented use in clinical settings.

\section{Methodology}
\subsection{Study Design}
We conducted a prospective
mixed-methods exploratory feasibility pilot study to evaluate the acceptability, perceived safety, and usability of the Molhim platform for PTSD screening in a military healthcare context.
The study employed both quantitative and qualitative data collection methods, including Likert-scale surveys, clinician reflection forms, and open-ended feedback. The study protocol was reviewed and approved by a military healthcare Institutional Review Board (IRB) and conducted in accordance with the principles of the Declaration of Helsinki. All participants provided informed consent prior to participation. Given the small sample size, the study is intended to provide preliminary insights into feasibility and acceptability rather than generalizable conclusions.

Given the early-stage nature of the system and the sensitive context of trauma-related screening, the study was designed as an observer-based evaluation rather than a direct interaction study. This approach was selected to minimize potential psychological risk, ensure clinician oversight, and allow safe initial assessment of system behavior before deployment with active users. As such, participants observed therapist-guided demonstrations of the system rather than engaging directly with it.

\subsection{Setting and Participants}
The study was conducted at a Ministry of Defense Health Services center in Riyadh, Saudi Arabia. A total of ten soldiers participated as observer-participants, recruited from the facility according to predefined inclusion criteria. Eligible participants were adults aged 21 years or older, in accordance with institutional recruitment criteria, with returning soldier status and the ability to complete a brief post-session survey. A licensed clinical psychologist conducted the demonstration sessions and provided observational reflections after each session.

\subsection{Procedure}
Each session consisted of a supervised demonstration by a licensed therapist of Molhim operating in PTSD screening mode. To minimize psychological risk during early-stage testing, soldiers participated as observers rather than direct users. Following each session, soldiers completed an acceptability survey assessing their perceptions of the system, and the therapist completed an observational reflection form documenting participant reactions, system behavior, and any safety concerns. Participants were debriefed after each session to ensure that no distress occurred.

\subsection{Measures}
Data collection combined quantitative ratings with qualitative feedback to capture both acceptability and safety perceptions. A study-specific instrument was developed for early-stage evaluation.

\begin{itemize}

\item \textbf{Soldier Acceptability Survey:} Soldiers completed a six-item Likert-scale questionnaire (1 = strongly disagree, 5 = strongly agree) assessing comfort observing the session, perceived clarity and appropriateness of the avatar's communication, perceived educational value regarding PTSD, perceived safety and non-judgment, perceived usefulness for other soldiers, and support for continued development of the system. Soldiers also responded to open-ended questions regarding what they liked most about the tool and suggested improvements. A binary safety check item asked whether any part of the session caused discomfort.

\item \textbf{Clinician Reflection Form:} After each session, the therapist completed a four-item binary checklist documenting whether the soldier appeared comfortable, whether signs of distress were observed, whether any inappropriate or mismatched system responses occurred, and whether the system appeared helpful for soldiers. The clinician also provided open-ended qualitative reflections describing usability observations, engagement, and suggestions for improvement.

\item \textbf{Qualitative Feedback:} Open-ended responses from both soldiers and the clinician were analyzed to identify recurring themes related to usability, communication quality, cultural alignment, and potential areas for system improvement.

\end{itemize}

\section{Results}

\subsection{Quantitative Findings: Soldier Acceptability Survey}
Overall ratings were favorable, with an average score of \textbf{4.12} across all survey items, indicating generally positive perceptions (above the neutral midpoint of 3 on a 5-point scale). Table~\ref{tab:soldier_likert} summarizes the mean and standard deviation for each survey question. The strongest endorsement was observed for continued development of the system (\textbf{4.50}). Perceived respectfulness of the avatar’s communication, educational value of the session, and perceived safety each received mean ratings of \textbf{4.20}. Comfort observing the session and belief that other soldiers may find the system helpful both received mean ratings of \textbf{3.80}. Scores around 4.20 reflect consistently positive perceptions, while scores of 3.80 suggest moderate agreement with slightly greater variability across participants. Standard deviations were generally low (overall SD = 0.71), indicating relatively consistent responses across participants.

\begin{table}[h]
\centering
\caption{Soldier Acceptability Survey Results (1–5)}
\label{tab:soldier_likert}
\rowcolors{2}{gray!20}{white}
\begin{tabular}{p{5.5cm}cc}
\hline
\textbf{Survey Item} & \textbf{Mean} & \textbf{SD} \\
\hline
I felt comfortable observing the session & 3.80 & 0.87 \\
The avatar's communication felt respectful and appropriate & 4.20 & 0.60 \\
The session helped me understand PTSD a little better & 4.20 & 0.60 \\
The system felt safe and non-judgmental & 4.20 & 0.40 \\
I believe some soldiers might find this system helpful & 3.80 & 0.87 \\
I support continuing development of this tool & \textbf{4.50} & 0.50 \\
\hline
\end{tabular}
\end{table}

A safety check item asked participants whether any part of the session caused discomfort. Eight of ten soldiers (\textbf{80\%}) reported no discomfort, while two participants (\textbf{20\%}) reported experiencing some level of discomfort during the demonstration.

\subsection{Clinician Observational Reflections}
Following each demonstration session, the supervising clinician completed a structured reflection form assessing participant reactions and system behavior. Table~\ref{tab:clinician_reflection} summarizes the results. Clinician observations indicated that the majority of soldiers appeared comfortable during the session (\textbf{90\%}). Signs of distress were observed in two sessions (\textbf{20\%}). Clinicians reported no inappropriate or mismatched system responses across all demonstrations. The clinician also indicated in every session that the tool appeared potentially helpful for soldiers (\textbf{10 out of 10} sessions).

\begin{table}[h]
\centering
\caption{Clinician Reflection Form Results}
\label{tab:clinician_reflection}
\rowcolors{2}{gray!20}{white}
\begin{tabular}{p{6cm}cc}
\hline
\textbf{Reflection Item} & \textbf{Yes} & \textbf{No} \\
\hline
Soldier appeared comfortable & \textbf{9} & 1 \\
Signs of distress observed & 2 & \textbf{8} \\
Inappropriate or mismatched responses & 0 & \textbf{10} \\
System may be helpful to soldiers & \textbf{10} & 0 \\
\hline
\end{tabular}
\end{table}

\subsection{Qualitative Findings: Open-Ended Responses}
\textbf{Perceived Strengths:} Several participants highlighted the system's ease of use and accessibility. Multiple soldiers reported that the interaction helped them better understand PTSD symptoms and appreciated the ability to ask questions through a structured conversational format. Participants also noted the potential value of the tool for supporting soldiers and providing quick post-session feedback and analysis.

\noindent\textbf{Suggested Improvements:} The most frequently suggested areas for improvement were response latency and interaction speed and improving TTS quality for the Saudi dialect. Some participants also suggested expanding functionality with additional educational materials, coping strategies, or interactive features.

\section{Discussion}
\subsection{Feasibility and Acceptability}
The findings of this pilot study provide preliminary evidence that a multimodal conversational virtual agent may be feasible for deployment for PTSD screening support. Overall soldier responses to the demonstration were favorable, with average Likert ratings above four across most survey dimensions and strong support for continued development of the system. Participants also largely reported feeling comfortable observing the interaction and described the avatar’s communication style as respectful and non-judgmental.

These results are consistent with prior research demonstrating that conversational agents can serve as acceptable interfaces for mental health screening and psychoeducation \cite{han2023preliminary,han2024assessing}. Studies of virtual human interviewers have similarly shown that individuals are willing to engage with AI-based systems in discussions of sensitive psychological topics \cite{han2021ptsdialogue, han2023preliminary, han2024assessing}. Our findings extend this line of work by suggesting that a culturally adapted Arabic-language conversational agent may be feasible for introduction into a military healthcare environment with positive perceptions of safety, respectfulness, and potential clinical usefulness. Clinicians reported no inappropriate or mismatched responses during any of the sessions, suggesting that the conversational architecture successfully maintained clinically appropriate dialogue behavior during the demonstration scenarios. While the system remains an early-stage prototype, the absence of critical safety failures during testing provides encouraging evidence for the viability of structured conversational protocols in mental health–related virtual agents.

\subsection{Social Cognition and the Disclosure Advantage}
One of the central motivations for using virtual agents in mental health screening is the disclosure advantage observed in human–computer interaction. Prior studies have shown that individuals may feel more comfortable sharing sensitive information with automated systems than with human clinicians, particularly when the system is perceived as non-judgmental and autonomous \cite{lucas2014s}.

Although participants in the present study did not interact directly with the system, their responses suggest that the design of the Molhim avatar aligns with several of the social factors known to facilitate disclosure. Participants rated the system highly in terms of respectful communication and perceived safety, both of which are critical elements in fostering trust during mental health conversations. These characteristics may reduce perceived social risk when discussing psychological distress, a particularly important factor in military populations where stigma and concerns about professional consequences can inhibit help-seeking behavior.

Social interaction cues such as conversational grounding and consistent turn-taking play a central role in establishing mutual understanding in dialogue \cite{clark1991grounding}, and have been shown to influence the perceived naturalness and coordination of interaction in conversational systems \cite{skantze2021turn}. The integration of a conversational avatar with speech-based interaction may therefore help bridge the gap between purely text-based chatbots and more socially expressive virtual interviewers, potentially strengthening user engagement and openness during screening interactions.

\subsection{Human-AI Cooperation in Clinical Virtual Agents}
Rather than positioning AI systems as replacements for clinicians, the system is designed to function within a cooperative human–AI workflow for mental health screening. In the pilot study, demonstration sessions were conducted under the supervision of a licensed clinician. This design choice was made intentionally to ensure participant safety and to minimize the risk of triggering distress during early-stage evaluation of the system.

The platform is intended to support self-directed screening interactions in which individuals engage with the virtual agent privately. During these interactions, the system can administer screening instruments, provide grounding exercises, and offer psychoeducational information. When elevated risk indicators are detected or users express a need for additional support, the system is designed to encourage referral to qualified mental health professionals.

Such architectures allow conversational agents to handle structured screening tasks and initial information gathering while clinicians remain responsible for diagnosis, treatment planning, and therapeutic intervention. This division of roles reflects a broader trend in clinical AI systems in which automated tools augment clinical workflows rather than replace human expertise \cite{topol2019high}.

In military healthcare settings, where stigma and workload constraints can limit access to early screening, self-directed conversational systems may offer a low-barrier entry point for individuals who might otherwise avoid discussing mental health concerns. By combining private screening interactions with pathways to professional care when needed, cooperative human–AI systems have the potential to expand access to early mental health assessment while preserving the central role of trained clinicians in care delivery.

\subsection{Cultural and Linguistic Design Implications}
Participant feedback highlighted language naturalness and interaction speed as key factors for deploying conversational AI in culturally specific contexts. Several participants suggested improvements to the Arabic speech quality and dialectal alignment of the avatar, indicating that linguistic authenticity plays an important role in perceived credibility and comfort. These observations align with broader research on AI adoption in healthcare, which shows that cultural and linguistic alignment strongly influence trust in automated systems. In Arabic-speaking regions, dialect variation, communication style, and culturally appropriate phrasing can significantly affect system performance, highlighting limitations in real-world deployment when models lack such alignment \cite{al2026words}. Response latency was also identified as a usability concern, as delays between user input and system response can disrupt conversational flow and reduce perceived social presence \cite{gnewuch2022opposing}. Addressing these challenges through improved dialect support, optimized speech processing, and efficient model inference will be important for enhancing interaction quality in future iterations of the system.

\subsection{Limitations}
Several limitations should be considered when interpreting the results of this study. First, the limited sample size of ten observer-participants restricts the statistical generalizability of the findings. The study involved sensitive clinical contexts and access to military healthcare environments, where participant recruitment and data collection are necessarily limited due to privacy, ethical, and institutional constraints. Second, participants observed therapist-led demonstrations rather than interacting directly with the system. While this design minimized potential psychological risk during early-stage testing, it does not capture the full range of user experiences that may occur during direct interaction. Because the study focused primarily on acceptability and perceived safety rather than clinical effectiveness, no diagnostic outcomes or symptom change measures were collected and this study cannot determine whether the system improves PTSD detection or mental health outcomes. Additionally, the visual component relies on a single-frame-per-turn sampling strategy and was not independently validated for affect recognition; therefore, visual signals are used only as lightweight contextual cues, and future work is needed to evaluate their reliability and contribution to screening performance.

\subsection{Future Work}
Future development of the platform will focus on improving conversational naturalness, expanding Arabic dialect support, and reducing system response latency. Further research into additional multimodal sensing capabilities could enable robust detection of behavioral indicators associated with psychological distress. Future studies should examine direct soldier interaction with the system and compare conversational screening performance with clinician-administered assessments. Larger-scale deployments across military healthcare facilities could also provide insight into how virtual agents integrate into existing clinical workflows and whether they meaningfully increase access to early mental health screening.

\section{Conclusion}
This paper presents Molhim, a multimodal conversational AI platform for purpose-specific interactions. We introduce a PTSD screening configuration for military healthcare that supports structured, goal-oriented dialogue aligned with clinical screening practices. Results from our mixed-methods feasibility pilot suggest generally positive perceptions, with high ratings for respectful communication, educational value, and perceived safety, and clinician observations confirming consistent potential usefulness for soldiers. This work further contributes to conversational AI research by demonstrating how multimodal systems can operate within culturally specific healthcare environments while supporting human–AI workflows. By continuing to explore socially intelligent and culturally adapted AI systems, we aim to advance the role of virtual agents as scalable tools for early mental health support and screening in diverse clinical settings.

\section{Acknowledgments}
The authors would like to thank Sama Almuraykhi and Sara Alghamdi from the Ministry of Defense Digital Transformation, Saudi Arabia, for their support in facilitating coordination and enabling deployment, as well as Dr. Youssef Alshehri, Dr. Waleed Alhazzani, Dr. Hadeel Abdulmohsen AlKhamees, and Dr. Mohammed Alnoyser for their guidance and continued support throughout this work.

\bibliographystyle{ACM-Reference-Format}
\bibliography{references}


\begin{thebibliography}{46}


\ifx \showCODEN    \undefined \def \showCODEN     #1{\unskip}     \fi
\ifx \showISBNx    \undefined \def \showISBNx     #1{\unskip}     \fi
\ifx \showISBNxiii \undefined \def \showISBNxiii  #1{\unskip}     \fi
\ifx \showISSN     \undefined \def \showISSN      #1{\unskip}     \fi
\ifx \showLCCN     \undefined \def \showLCCN      #1{\unskip}     \fi
\ifx \shownote     \undefined \def \shownote      #1{#1}          \fi
\ifx \showarticletitle \undefined \def \showarticletitle #1{#1}   \fi
\ifx \showURL      \undefined \def \showURL       {\relax}        \fi
\providecommand\bibfield[2]{#2}
\providecommand\bibinfo[2]{#2}
\providecommand\natexlab[1]{#1}
\providecommand\showeprint[2][]{arXiv:#2}

\bibitem[Al-Ghadhban and Al-Twairesh(2020)]%
        {al2020nabiha}
\bibfield{author}{\bibinfo{person}{Dana Al-Ghadhban} {and} \bibinfo{person}{Nora Al-Twairesh}.} \bibinfo{year}{2020}\natexlab{}.
\newblock \showarticletitle{Nabiha: an Arabic dialect chatbot}.
\newblock \bibinfo{journal}{\emph{International Journal of Advanced Computer Science and Applications}} \bibinfo{volume}{11}, \bibinfo{number}{3} (\bibinfo{year}{2020}).
\newblock


\bibitem[Al~Hanai et~al\mbox{.}(2018)]%
        {al2018detecting}
\bibfield{author}{\bibinfo{person}{Tuka Al~Hanai}, \bibinfo{person}{Mohammad~M Ghassemi}, {and} \bibinfo{person}{James~R Glass}.} \bibinfo{year}{2018}\natexlab{}.
\newblock \showarticletitle{Detecting depression with audio/text sequence modeling of interviews.}. In \bibinfo{booktitle}{\emph{Interspeech}}. \bibinfo{pages}{1716--1720}.
\newblock


\bibitem[Al-Monef et~al\mbox{.}(2026)]%
        {al2026words}
\bibfield{author}{\bibinfo{person}{Renad Al-Monef}, \bibinfo{person}{Hassan Alhuzali}, \bibinfo{person}{Nora Alturayeif}, {and} \bibinfo{person}{Ashwag Alasmari}.} \bibinfo{year}{2026}\natexlab{}.
\newblock \showarticletitle{From words to proverbs: Evaluating LLMs’ linguistic and cultural competence in Saudi dialects with Absher}.
\newblock \bibinfo{journal}{\emph{Alexandria Engineering Journal}}  \bibinfo{volume}{137} (\bibinfo{year}{2026}), \bibinfo{pages}{25--41}.
\newblock


\bibitem[Alghamdi et~al\mbox{.}(2025)]%
        {alghamdi2025effect}
\bibfield{author}{\bibinfo{person}{Essam Alghamdi}, \bibinfo{person}{Martin Halvey}, {and} \bibinfo{person}{Emma Nicol}.} \bibinfo{year}{2025}\natexlab{}.
\newblock \showarticletitle{The effect of interaction language on preferences for communication repair strategies in Digital Voice Assistants (DVAs): a comparative study}.
\newblock In \bibinfo{booktitle}{\emph{Proceedings of the Mensch und Computer 2025}}. \bibinfo{pages}{146--162}.
\newblock


\bibitem[Alharbi et~al\mbox{.}(2026)]%
        {alharbi2026determinants}
\bibfield{author}{\bibinfo{person}{Bandar~S Alharbi}, \bibinfo{person}{Majed~M Aljabri}, {and} \bibinfo{person}{Endale~Alemayehu Ali}.} \bibinfo{year}{2026}\natexlab{}.
\newblock \showarticletitle{Determinants of Trust in Artificial Intelligence (AI) for Health-Related Decision-Making Among Adults in Saudi Arabia: A Cross-Sectional Study}. In \bibinfo{booktitle}{\emph{Healthcare}}, Vol.~\bibinfo{volume}{14}. MDPI, \bibinfo{pages}{506}.
\newblock


\bibitem[Ali and Hoque(2017)]%
        {ali2017social}
\bibfield{author}{\bibinfo{person}{Mohammad~Rafayet Ali} {and} \bibinfo{person}{Ehsan Hoque}.} \bibinfo{year}{2017}\natexlab{}.
\newblock \showarticletitle{Social skills training with virtual assistant and real-time feedback}. In \bibinfo{booktitle}{\emph{Proceedings of the 2017 ACM international joint conference on pervasive and ubiquitous computing and proceedings of the 2017 ACM international symposium on wearable computers}}. \bibinfo{pages}{325--329}.
\newblock


\bibitem[Armenta et~al\mbox{.}(2018)]%
        {armenta2018factors}
\bibfield{author}{\bibinfo{person}{Richard~F Armenta}, \bibinfo{person}{Toni Rush}, \bibinfo{person}{Cynthia~A LeardMann}, \bibinfo{person}{Jeffrey Millegan}, \bibinfo{person}{Adam Cooper}, \bibinfo{person}{Charles~W Hoge}, {and} \bibinfo{person}{Millennium Cohort~Study team}.} \bibinfo{year}{2018}\natexlab{}.
\newblock \showarticletitle{Factors associated with persistent posttraumatic stress disorder among US military service members and veterans}.
\newblock \bibinfo{journal}{\emph{BMC psychiatry}} \bibinfo{volume}{18}, \bibinfo{number}{1} (\bibinfo{year}{2018}), \bibinfo{pages}{48}.
\newblock


\bibitem[Bliese et~al\mbox{.}(2008)]%
        {bliese2008validating}
\bibfield{author}{\bibinfo{person}{Paul~D. Bliese}, \bibinfo{person}{Kathleen~M. Wright}, \bibinfo{person}{Amy~B. Adler}, \bibinfo{person}{Oscar Cabrera}, \bibinfo{person}{Carl~A. Castro}, {and} \bibinfo{person}{Charles~W. Hoge}.} \bibinfo{year}{2008}\natexlab{}.
\newblock \showarticletitle{Validating the primary care posttraumatic stress disorder screen and the posttraumatic stress disorder checklist with soldiers returning from combat}.
\newblock \bibinfo{journal}{\emph{Journal of Consulting and Clinical Psychology}} \bibinfo{volume}{76}, \bibinfo{number}{2} (\bibinfo{year}{2008}), \bibinfo{pages}{272--281}.
\newblock
\href{https://doi.org/10.1037/0022-006X.76.2.272}{doi:\nolinkurl{10.1037/0022-006X.76.2.272}}


\bibitem[Chen et~al\mbox{.}(2025)]%
        {chen2025detecting}
\bibfield{author}{\bibinfo{person}{Feng Chen}, \bibinfo{person}{Dror Ben-Zeev}, \bibinfo{person}{Gillian Sparks}, \bibinfo{person}{Arya Kadakia}, {and} \bibinfo{person}{Trevor Cohen}.} \bibinfo{year}{2025}\natexlab{}.
\newblock \showarticletitle{Detecting PTSD in clinical interviews: A comparative analysis of NLP methods and large language models}. In \bibinfo{booktitle}{\emph{Biocomputing 2026: Proceedings of the Pacific Symposium}}. World Scientific, \bibinfo{pages}{265--279}.
\newblock


\bibitem[Clark and Brennan(1991)]%
        {clark1991grounding}
\bibfield{author}{\bibinfo{person}{Herbert~H. Clark} {and} \bibinfo{person}{Susan~E. Brennan}.} \bibinfo{year}{1991}\natexlab{}.
\newblock \showarticletitle{Grounding in Communication}.
\newblock In \bibinfo{booktitle}{\emph{Perspectives on Socially Shared Cognition}}, \bibfield{editor}{\bibinfo{person}{Lauren~B. Resnick}, \bibinfo{person}{John~M. Levine}, {and} \bibinfo{person}{Stephanie~D. Teasley}} (Eds.). \bibinfo{publisher}{American Psychological Association}, \bibinfo{pages}{127--149}.
\newblock
\href{https://doi.org/10.1037/10096-006}{doi:\nolinkurl{10.1037/10096-006}}


\bibitem[Croes et~al\mbox{.}(2024)]%
        {croes2024digital}
\bibfield{author}{\bibinfo{person}{Emmelyn~AJ Croes}, \bibinfo{person}{Marjolijn~L Antheunis}, \bibinfo{person}{Chris Van Der~Lee}, {and} \bibinfo{person}{Jan~MS De~Wit}.} \bibinfo{year}{2024}\natexlab{}.
\newblock \showarticletitle{Digital confessions: The willingness to disclose intimate information to a chatbot and its impact on emotional well-being}.
\newblock \bibinfo{journal}{\emph{Interacting with Computers}} \bibinfo{volume}{36}, \bibinfo{number}{5} (\bibinfo{year}{2024}), \bibinfo{pages}{279--292}.
\newblock


\bibitem[DeVault et~al\mbox{.}(2014)]%
        {devault2014simsensei}
\bibfield{author}{\bibinfo{person}{David DeVault}, \bibinfo{person}{Ron Artstein}, \bibinfo{person}{Grace Benn}, \bibinfo{person}{Teresa Dey}, \bibinfo{person}{Ed Fast}, \bibinfo{person}{Alesia Gainer}, \bibinfo{person}{Kallirroi Georgila}, \bibinfo{person}{Jon Gratch}, \bibinfo{person}{Arno Hartholt}, \bibinfo{person}{Margaux Lhommet}, {et~al\mbox{.}}} \bibinfo{year}{2014}\natexlab{}.
\newblock \showarticletitle{SimSensei Kiosk: A virtual human interviewer for healthcare decision support}. In \bibinfo{booktitle}{\emph{Proceedings of the 2014 international conference on Autonomous agents and multi-agent systems}}. \bibinfo{pages}{1061--1068}.
\newblock


\bibitem[Fadhil et~al\mbox{.}(2019)]%
        {fadhil2019ollobot}
\bibfield{author}{\bibinfo{person}{Ahmed Fadhil} {et~al\mbox{.}}} \bibinfo{year}{2019}\natexlab{}.
\newblock \showarticletitle{OlloBot-towards a text-based arabic health conversational agent: Evaluation and results}. In \bibinfo{booktitle}{\emph{Proceedings of the international conference on recent advances in natural language processing (RANLP 2019)}}. \bibinfo{pages}{295--303}.
\newblock


\bibitem[Farghaly and Shaalan(2009)]%
        {farghaly2009arabic}
\bibfield{author}{\bibinfo{person}{Ali Farghaly} {and} \bibinfo{person}{Khaled Shaalan}.} \bibinfo{year}{2009}\natexlab{}.
\newblock \showarticletitle{Arabic natural language processing: Challenges and solutions}.
\newblock \bibinfo{journal}{\emph{ACM Transactions on Asian Language Information Processing (TALIP)}} \bibinfo{volume}{8}, \bibinfo{number}{4} (\bibinfo{year}{2009}), \bibinfo{pages}{1--22}.
\newblock


\bibitem[Fitzpatrick et~al\mbox{.}(2017)]%
        {fitzpatrick2017delivering}
\bibfield{author}{\bibinfo{person}{Kathleen~Kara Fitzpatrick}, \bibinfo{person}{Alison Darcy}, {and} \bibinfo{person}{Molly Vierhile}.} \bibinfo{year}{2017}\natexlab{}.
\newblock \showarticletitle{Delivering cognitive behavior therapy to young adults with symptoms of depression and anxiety using a fully automated conversational agent (Woebot): a randomized controlled trial}.
\newblock \bibinfo{journal}{\emph{JMIR mental health}} \bibinfo{volume}{4}, \bibinfo{number}{2} (\bibinfo{year}{2017}), \bibinfo{pages}{e7785}.
\newblock


\bibitem[Fulmer et~al\mbox{.}(2018)]%
        {fulmer2018using}
\bibfield{author}{\bibinfo{person}{Russell Fulmer}, \bibinfo{person}{Angela Joerin}, \bibinfo{person}{Breanna Gentile}, \bibinfo{person}{Lysanne Lakerink}, {and} \bibinfo{person}{Michiel Rauws}.} \bibinfo{year}{2018}\natexlab{}.
\newblock \showarticletitle{Using psychological artificial intelligence (Tess) to relieve symptoms of depression and anxiety: randomized controlled trial}.
\newblock \bibinfo{journal}{\emph{JMIR mental health}} \bibinfo{volume}{5}, \bibinfo{number}{4} (\bibinfo{year}{2018}), \bibinfo{pages}{e9782}.
\newblock


\bibitem[Gnewuch et~al\mbox{.}(2022)]%
        {gnewuch2022opposing}
\bibfield{author}{\bibinfo{person}{Ulrich Gnewuch}, \bibinfo{person}{Stefan Morana}, \bibinfo{person}{Marc~TP Adam}, {and} \bibinfo{person}{Alexander Maedche}.} \bibinfo{year}{2022}\natexlab{}.
\newblock \showarticletitle{Opposing Effects of Response Time in Human--Chatbot Interaction: U. Gnewuch et al.}
\newblock \bibinfo{journal}{\emph{Business \& Information Systems Engineering}} \bibinfo{volume}{64}, \bibinfo{number}{6} (\bibinfo{year}{2022}), \bibinfo{pages}{773--791}.
\newblock


\bibitem[Gratch et~al\mbox{.}(2014)]%
        {gratch2014daic}
\bibfield{author}{\bibinfo{person}{Jonathan Gratch}, \bibinfo{person}{Ron Artstein}, \bibinfo{person}{Gale Lucas}, {and} \bibinfo{person}{et al.}} \bibinfo{year}{2014}\natexlab{}.
\newblock \showarticletitle{The Distress Analysis Interview Corpus of Human and Computer Interviews}. In \bibinfo{booktitle}{\emph{LREC}}.
\newblock


\bibitem[Han et~al\mbox{.}(2021)]%
        {han2021ptsdialogue}
\bibfield{author}{\bibinfo{person}{Hee~Jeong Han}, \bibinfo{person}{Sanjana Mendu}, \bibinfo{person}{Beth~K Jaworski}, \bibinfo{person}{Jason E~Owen}, {and} \bibinfo{person}{Saeed Abdullah}.} \bibinfo{year}{2021}\natexlab{}.
\newblock \showarticletitle{PTSDialogue: designing a conversational agent to support individuals with post-traumatic stress disorder}. In \bibinfo{booktitle}{\emph{Adjunct proceedings of the 2021 ACM international joint conference on pervasive and ubiquitous computing and proceedings of the 2021 ACM international symposium on wearable computers}}. \bibinfo{pages}{198--203}.
\newblock


\bibitem[Han et~al\mbox{.}(2023)]%
        {han2023preliminary}
\bibfield{author}{\bibinfo{person}{Hee~Jeong Han}, \bibinfo{person}{Sanjana Mendu}, \bibinfo{person}{Beth~K Jaworski}, \bibinfo{person}{Jason~E Owen}, {and} \bibinfo{person}{Saeed Abdullah}.} \bibinfo{year}{2023}\natexlab{}.
\newblock \showarticletitle{Preliminary evaluation of a conversational agent to support self-management of individuals living with posttraumatic stress disorder: interview study with clinical experts}.
\newblock \bibinfo{journal}{\emph{JMIR Formative Research}}  \bibinfo{volume}{7} (\bibinfo{year}{2023}), \bibinfo{pages}{e45894}.
\newblock


\bibitem[Han et~al\mbox{.}(2024)]%
        {han2024assessing}
\bibfield{author}{\bibinfo{person}{Hee~Jeong Han}, \bibinfo{person}{Sanjana Mendu}, \bibinfo{person}{Beth~K Jaworski}, \bibinfo{person}{Jason~E Owen}, {and} \bibinfo{person}{Saeed Abdullah}.} \bibinfo{year}{2024}\natexlab{}.
\newblock \showarticletitle{Assessing acceptance and feasibility of a conversational agent to support individuals living with post-traumatic stress disorder}.
\newblock \bibinfo{journal}{\emph{Digital Health}}  \bibinfo{volume}{10} (\bibinfo{year}{2024}), \bibinfo{pages}{20552076241286133}.
\newblock


\bibitem[Hasan et~al\mbox{.}(2023)]%
        {hasan2023sapien}
\bibfield{author}{\bibinfo{person}{Masum Hasan}, \bibinfo{person}{Cengiz Ozel}, \bibinfo{person}{Sammy Potter}, {and} \bibinfo{person}{Ehsan Hoque}.} \bibinfo{year}{2023}\natexlab{}.
\newblock \showarticletitle{SAPIEN: affective virtual agents powered by large language models}. In \bibinfo{booktitle}{\emph{2023 11th international conference on affective computing and intelligent interaction workshops and demos (ACIIW)}}. IEEE, \bibinfo{pages}{1--3}.
\newblock


\bibitem[Hoge et~al\mbox{.}(2004)]%
        {hoge2004combat}
\bibfield{author}{\bibinfo{person}{Charles~W Hoge}, \bibinfo{person}{Carl~A Castro}, \bibinfo{person}{Stephen~C Messer}, \bibinfo{person}{Dennis McGurk}, \bibinfo{person}{Dave~I Cotting}, {and} \bibinfo{person}{Robert~L Koffman}.} \bibinfo{year}{2004}\natexlab{}.
\newblock \showarticletitle{Combat duty in Iraq and Afghanistan, mental health problems, and barriers to care}.
\newblock \bibinfo{journal}{\emph{New England journal of medicine}} \bibinfo{volume}{351}, \bibinfo{number}{1} (\bibinfo{year}{2004}), \bibinfo{pages}{13--22}.
\newblock


\bibitem[Hu et~al\mbox{.}(2024)]%
        {hu2024speech}
\bibfield{author}{\bibinfo{person}{Jiawei Hu}, \bibinfo{person}{Chunxiao Zhao}, \bibinfo{person}{Congrong Shi}, \bibinfo{person}{Ziyi Zhao}, {and} \bibinfo{person}{Zhihong Ren}.} \bibinfo{year}{2024}\natexlab{}.
\newblock \showarticletitle{Speech-based recognition and estimating severity of PTSD using machine learning}.
\newblock \bibinfo{journal}{\emph{Journal of affective disorders}}  \bibinfo{volume}{362} (\bibinfo{year}{2024}), \bibinfo{pages}{859--868}.
\newblock


\bibitem[Kane et~al\mbox{.}(2022)]%
        {kane2022flexible}
\bibfield{author}{\bibinfo{person}{Benjamin Kane}, \bibinfo{person}{Catherine Giugno}, \bibinfo{person}{Lenhart Schubert}, \bibinfo{person}{Kurtis Haut}, \bibinfo{person}{Caleb Wohn}, {and} \bibinfo{person}{Ehsan Hoque}.} \bibinfo{year}{2022}\natexlab{}.
\newblock \showarticletitle{A flexible schema-guided dialogue management framework: From friendly peer to virtual standardized cancer patient}.
\newblock \bibinfo{journal}{\emph{arXiv preprint arXiv:2207.07276}} (\bibinfo{year}{2022}).
\newblock


\bibitem[Kuhn et~al\mbox{.}(2014)]%
        {kuhn2014preliminary}
\bibfield{author}{\bibinfo{person}{Eric Kuhn}, \bibinfo{person}{Carolyn Greene}, \bibinfo{person}{Julia Hoffman}, \bibinfo{person}{Tam Nguyen}, \bibinfo{person}{Laura Wald}, \bibinfo{person}{Janet Schmidt}, \bibinfo{person}{Kelly~M Ramsey}, {and} \bibinfo{person}{Josef Ruzek}.} \bibinfo{year}{2014}\natexlab{}.
\newblock \showarticletitle{Preliminary evaluation of PTSD Coach, a smartphone app for post-traumatic stress symptoms}.
\newblock \bibinfo{journal}{\emph{Military medicine}} \bibinfo{volume}{179}, \bibinfo{number}{1} (\bibinfo{year}{2014}), \bibinfo{pages}{12--18}.
\newblock


\bibitem[Lee et~al\mbox{.}(2025)]%
        {lee2025artificial}
\bibfield{author}{\bibinfo{person}{Hyein~S Lee}, \bibinfo{person}{Colton Wright}, \bibinfo{person}{Julia Ferranto}, \bibinfo{person}{Jessica Buttimer}, \bibinfo{person}{Clare~E Palmer}, \bibinfo{person}{Andrew Welchman}, \bibinfo{person}{Kathleen~M Mazor}, \bibinfo{person}{Kimberly~A Fisher}, \bibinfo{person}{David Smelson}, \bibinfo{person}{Laurel O’Connor}, {et~al\mbox{.}}} \bibinfo{year}{2025}\natexlab{}.
\newblock \showarticletitle{Artificial intelligence conversational agents in mental health: Patients see potential, but prefer humans in the loop}.
\newblock \bibinfo{journal}{\emph{Frontiers in Psychiatry}}  \bibinfo{volume}{15} (\bibinfo{year}{2025}), \bibinfo{pages}{1505024}.
\newblock


\bibitem[Lucas et~al\mbox{.}(2014)]%
        {lucas2014s}
\bibfield{author}{\bibinfo{person}{Gale~M Lucas}, \bibinfo{person}{Jonathan Gratch}, \bibinfo{person}{Aisha King}, {and} \bibinfo{person}{Louis-Philippe Morency}.} \bibinfo{year}{2014}\natexlab{}.
\newblock \showarticletitle{It’s only a computer: Virtual humans increase willingness to disclose}.
\newblock \bibinfo{journal}{\emph{Computers in Human Behavior}}  \bibinfo{volume}{37} (\bibinfo{year}{2014}), \bibinfo{pages}{94--100}.
\newblock


\bibitem[Lucas et~al\mbox{.}(2017)]%
        {lucas2017reporting}
\bibfield{author}{\bibinfo{person}{Gale~M Lucas}, \bibinfo{person}{Albert Rizzo}, \bibinfo{person}{Jonathan Gratch}, \bibinfo{person}{Stefan Scherer}, \bibinfo{person}{Giota Stratou}, \bibinfo{person}{Jill Boberg}, {and} \bibinfo{person}{Louis-Philippe Morency}.} \bibinfo{year}{2017}\natexlab{}.
\newblock \showarticletitle{Reporting mental health symptoms: breaking down barriers to care with virtual human interviewers}.
\newblock \bibinfo{journal}{\emph{Frontiers in Robotics and AI}}  \bibinfo{volume}{4} (\bibinfo{year}{2017}), \bibinfo{pages}{51}.
\newblock


\bibitem[Megnin-Viggars et~al\mbox{.}(2019)]%
        {megnin2019post}
\bibfield{author}{\bibinfo{person}{Odette Megnin-Viggars}, \bibinfo{person}{Ifigeneia Mavranezouli}, \bibinfo{person}{Neil Greenberg}, \bibinfo{person}{Steve Hajioff}, {and} \bibinfo{person}{Jonathan Leach}.} \bibinfo{year}{2019}\natexlab{}.
\newblock \showarticletitle{Post-traumatic stress disorder: what does NICE guidance mean for primary care?}
\newblock \bibinfo{journal}{\emph{The British Journal of General Practice}} \bibinfo{volume}{69}, \bibinfo{number}{684} (\bibinfo{year}{2019}), \bibinfo{pages}{328}.
\newblock


\bibitem[Meng and Dai(2021)]%
        {meng2021emotional}
\bibfield{author}{\bibinfo{person}{Jingbo Meng} {and} \bibinfo{person}{Yue Dai}.} \bibinfo{year}{2021}\natexlab{}.
\newblock \showarticletitle{Emotional support from AI chatbots: Should a supportive partner self-disclose or not?}
\newblock \bibinfo{journal}{\emph{Journal of Computer-Mediated Communication}} \bibinfo{volume}{26}, \bibinfo{number}{4} (\bibinfo{year}{2021}), \bibinfo{pages}{207--222}.
\newblock


\bibitem[Milliken et~al\mbox{.}(2007)]%
        {milliken2007longitudinal}
\bibfield{author}{\bibinfo{person}{Charles~S Milliken}, \bibinfo{person}{Jennifer~L Auchterlonie}, {and} \bibinfo{person}{Charles~W Hoge}.} \bibinfo{year}{2007}\natexlab{}.
\newblock \showarticletitle{Longitudinal assessment of mental health problems among active and reserve component soldiers returning from the Iraq war}.
\newblock \bibinfo{journal}{\emph{Jama}} \bibinfo{volume}{298}, \bibinfo{number}{18} (\bibinfo{year}{2007}), \bibinfo{pages}{2141--2148}.
\newblock


\bibitem[Naderbagi et~al\mbox{.}(2024)]%
        {naderbagi2024cultural}
\bibfield{author}{\bibinfo{person}{Aila Naderbagi}, \bibinfo{person}{Victoria Loblay}, \bibinfo{person}{Iqthyer Uddin~Md Zahed}, \bibinfo{person}{Mahalakshmi Ekambareshwar}, \bibinfo{person}{Adam Poulsen}, \bibinfo{person}{Yun~JC Song}, \bibinfo{person}{Laura Ospina-Pinillos}, \bibinfo{person}{Michael Krausz}, \bibinfo{person}{Mostafa Mamdouh~Kamel}, \bibinfo{person}{Ian~B Hickie}, {et~al\mbox{.}}} \bibinfo{year}{2024}\natexlab{}.
\newblock \showarticletitle{Cultural and contextual adaptation of digital health interventions: narrative review}.
\newblock \bibinfo{journal}{\emph{Journal of medical Internet research}}  \bibinfo{volume}{26} (\bibinfo{year}{2024}), \bibinfo{pages}{e55130}.
\newblock


\bibitem[Owen et~al\mbox{.}(2015)]%
        {owen2015mhealth}
\bibfield{author}{\bibinfo{person}{Jason~E Owen}, \bibinfo{person}{Beth~K Jaworski}, \bibinfo{person}{Eric Kuhn}, \bibinfo{person}{Kerry~N Makin-Byrd}, \bibinfo{person}{Kelly~M Ramsey}, {and} \bibinfo{person}{Julia~E Hoffman}.} \bibinfo{year}{2015}\natexlab{}.
\newblock \showarticletitle{mHealth in the wild: using novel data to examine the reach, use, and impact of PTSD coach}.
\newblock \bibinfo{journal}{\emph{JMIR mental health}} \bibinfo{volume}{2}, \bibinfo{number}{1} (\bibinfo{year}{2015}), \bibinfo{pages}{e7}.
\newblock


\bibitem[Pelikan and Hofstetter(2023)]%
        {pelikan2023managing}
\bibfield{author}{\bibinfo{person}{Hannah Pelikan} {and} \bibinfo{person}{Emily Hofstetter}.} \bibinfo{year}{2023}\natexlab{}.
\newblock \showarticletitle{Managing delays in human-robot interaction}.
\newblock \bibinfo{journal}{\emph{ACM Transactions on Computer-Human Interaction}} \bibinfo{volume}{30}, \bibinfo{number}{4} (\bibinfo{year}{2023}), \bibinfo{pages}{1--42}.
\newblock


\bibitem[Pickard et~al\mbox{.}(2016)]%
        {pickard2016revealing}
\bibfield{author}{\bibinfo{person}{Matthew~D Pickard}, \bibinfo{person}{Catherine~A Roster}, {and} \bibinfo{person}{Yixing Chen}.} \bibinfo{year}{2016}\natexlab{}.
\newblock \showarticletitle{Revealing sensitive information in personal interviews: Is self-disclosure easier with humans or avatars and under what conditions?}
\newblock \bibinfo{journal}{\emph{Computers in Human Behavior}}  \bibinfo{volume}{65} (\bibinfo{year}{2016}), \bibinfo{pages}{23--30}.
\newblock


\bibitem[Polusny et~al\mbox{.}(2011)]%
        {polusny2011prospective}
\bibfield{author}{\bibinfo{person}{Melissa~A Polusny}, \bibinfo{person}{Chirstopher~R Erbes}, \bibinfo{person}{Maureen Murdoch}, \bibinfo{person}{Paul~A Arbisi}, \bibinfo{person}{Paul Thuras}, {and} \bibinfo{person}{MB Rath}.} \bibinfo{year}{2011}\natexlab{}.
\newblock \showarticletitle{Prospective risk factors for new-onset post-traumatic stress disorder in National Guard soldiers deployed to Iraq}.
\newblock \bibinfo{journal}{\emph{Psychological medicine}} \bibinfo{volume}{41}, \bibinfo{number}{4} (\bibinfo{year}{2011}), \bibinfo{pages}{687--698}.
\newblock


\bibitem[Rozek et~al\mbox{.}(2023)]%
        {rozek2023understanding}
\bibfield{author}{\bibinfo{person}{David~C Rozek}, \bibinfo{person}{Victoria~L Steigerwald}, \bibinfo{person}{Shelby~N Baker}, \bibinfo{person}{Georgina Gross}, \bibinfo{person}{Kelly~P Maieritsch}, \bibinfo{person}{Rani Hoff}, \bibinfo{person}{Ilan Harpaz-Rotem}, {and} \bibinfo{person}{Noelle~B Smith}.} \bibinfo{year}{2023}\natexlab{}.
\newblock \showarticletitle{Understanding veteran barriers to specialty outpatient PTSD clinical care}.
\newblock \bibinfo{journal}{\emph{Journal of anxiety disorders}}  \bibinfo{volume}{95} (\bibinfo{year}{2023}), \bibinfo{pages}{102675}.
\newblock


\bibitem[Sawalha et~al\mbox{.}(2022)]%
        {sawalha2022detecting}
\bibfield{author}{\bibinfo{person}{Jeff Sawalha}, \bibinfo{person}{Muhammad Yousefnezhad}, \bibinfo{person}{Zehra Shah}, \bibinfo{person}{Matthew~RG Brown}, \bibinfo{person}{Andrew~J Greenshaw}, {and} \bibinfo{person}{Russell Greiner}.} \bibinfo{year}{2022}\natexlab{}.
\newblock \showarticletitle{Detecting presence of PTSD using sentiment analysis from text data}.
\newblock \bibinfo{journal}{\emph{Frontiers in psychiatry}}  \bibinfo{volume}{12} (\bibinfo{year}{2022}), \bibinfo{pages}{811392}.
\newblock


\bibitem[Schuetzler et~al\mbox{.}(2018)]%
        {schuetzler2018influence}
\bibfield{author}{\bibinfo{person}{Ryan~M Schuetzler}, \bibinfo{person}{Justin~Scott Giboney}, \bibinfo{person}{G~Mark Grimes}, {and} \bibinfo{person}{Jay~F Nunamaker~Jr}.} \bibinfo{year}{2018}\natexlab{}.
\newblock \showarticletitle{The influence of conversational agent embodiment and conversational relevance on socially desirable responding}.
\newblock \bibinfo{journal}{\emph{Decision Support Systems}}  \bibinfo{volume}{114} (\bibinfo{year}{2018}), \bibinfo{pages}{94--102}.
\newblock


\bibitem[Sharp et~al\mbox{.}(2015)]%
        {sharp2015stigma}
\bibfield{author}{\bibinfo{person}{Marie-Louise Sharp}, \bibinfo{person}{Nicola~T Fear}, \bibinfo{person}{Roberto~J Rona}, \bibinfo{person}{Simon Wessely}, \bibinfo{person}{Neil Greenberg}, \bibinfo{person}{Norman Jones}, {and} \bibinfo{person}{Laura Goodwin}.} \bibinfo{year}{2015}\natexlab{}.
\newblock \showarticletitle{Stigma as a barrier to seeking health care among military personnel with mental health problems}.
\newblock \bibinfo{journal}{\emph{Epidemiologic reviews}} \bibinfo{volume}{37}, \bibinfo{number}{1} (\bibinfo{year}{2015}), \bibinfo{pages}{144--162}.
\newblock


\bibitem[Skantze(2021)]%
        {skantze2021turn}
\bibfield{author}{\bibinfo{person}{Gabriel Skantze}.} \bibinfo{year}{2021}\natexlab{}.
\newblock \showarticletitle{Turn-taking in conversational systems and human-robot interaction: a review}.
\newblock \bibinfo{journal}{\emph{Computer Speech \& Language}}  \bibinfo{volume}{67} (\bibinfo{year}{2021}), \bibinfo{pages}{101178}.
\newblock


\bibitem[Steenkamp et~al\mbox{.}(2015)]%
        {steenkamp2015psychotherapy}
\bibfield{author}{\bibinfo{person}{Maria~M Steenkamp}, \bibinfo{person}{Brett~T Litz}, \bibinfo{person}{Charles~W Hoge}, {and} \bibinfo{person}{Charles~R Marmar}.} \bibinfo{year}{2015}\natexlab{}.
\newblock \showarticletitle{Psychotherapy for military-related PTSD: A review of randomized clinical trials}.
\newblock \bibinfo{journal}{\emph{Jama}} \bibinfo{volume}{314}, \bibinfo{number}{5} (\bibinfo{year}{2015}), \bibinfo{pages}{489--500}.
\newblock


\bibitem[Tielman et~al\mbox{.}(2017)]%
        {tielman2017therapy}
\bibfield{author}{\bibinfo{person}{Myrthe~L Tielman}, \bibinfo{person}{Mark~A Neerincx}, \bibinfo{person}{Rafael Bidarra}, \bibinfo{person}{Ben Kybartas}, {and} \bibinfo{person}{Willem-Paul Brinkman}.} \bibinfo{year}{2017}\natexlab{}.
\newblock \showarticletitle{A therapy system for post-traumatic stress disorder using a virtual agent and virtual storytelling to reconstruct traumatic memories}.
\newblock \bibinfo{journal}{\emph{Journal of medical systems}} \bibinfo{volume}{41}, \bibinfo{number}{8} (\bibinfo{year}{2017}), \bibinfo{pages}{125}.
\newblock


\bibitem[Topol(2019)]%
        {topol2019high}
\bibfield{author}{\bibinfo{person}{Eric~J Topol}.} \bibinfo{year}{2019}\natexlab{}.
\newblock \showarticletitle{High-performance medicine: the convergence of human and artificial intelligence}.
\newblock \bibinfo{journal}{\emph{Nature medicine}} \bibinfo{volume}{25}, \bibinfo{number}{1} (\bibinfo{year}{2019}), \bibinfo{pages}{44--56}.
\newblock


\bibitem[Warner et~al\mbox{.}(2008)]%
        {warner2008soldier}
\bibfield{author}{\bibinfo{person}{Christopher~H Warner}, \bibinfo{person}{George~N Appenzeller}, \bibinfo{person}{Keri Mullen}, \bibinfo{person}{Carolynn~M Warner}, {and} \bibinfo{person}{Thomas Grieger}.} \bibinfo{year}{2008}\natexlab{}.
\newblock \showarticletitle{Soldier attitudes toward mental health screening and seeking care upon return from combat}.
\newblock \bibinfo{journal}{\emph{Military medicine}} \bibinfo{volume}{173}, \bibinfo{number}{6} (\bibinfo{year}{2008}), \bibinfo{pages}{563--569}.
\newblock


\end{thebibliography}

\end{document}